\documentclass[11pt]{article}
\usepackage{amssymb}
\usepackage{pifont}
\usepackage{amsmath}
\usepackage{epic,eepic}
\makeatletter%
\newcount\hour  \newcount\minutes  \hour=\time  \divide\hour by 60
\minutes=\hour  \multiply\minutes by -60  \advance\minutes by \time
\def\mmmddyyyy{\ifcase\month\or Jan\or Feb\or Mar\or Apr\or May\or Jun\or Jul\o\
r
  Aug\or Sep\or Oct\or Nov\or Dec\fi \space\number\day, \number\year}
\def\hhmm{\ifnum\hour<10 0\fi\number\hour :%
  \ifnum\minutes<10 0\fi\number\minutes}

\newcommand{\singlespacing}{\let\CS=
\@currsize\renewcommand{\baselinestretch}{1}\tiny\CS}
\newcommand{\singlespacingplus}{\let\CS=
\@currsize\renewcommand{\baselinestretch}{1.25}\tiny\CS}
\newcommand{\nicefivespacing}{\let\CS=
\@currsize\renewcommand{\baselinestretch}{1.5}\tiny\CS}
\newcommand{\doublespacing}{\let\CS=
\@currsize\renewcommand{\baselinestretch}{1.75}\tiny\CS}
\newcommand{\mydoublespacing}{\let\CS=
\@currsize\renewcommand{\baselinestretch}{1.75}\tiny\CS}
\newcommand{\draftspacing}{\let\CS=
\@currsize\renewcommand{\baselinestretch}{2.0}\tiny\CS}
\newcommand{\normalspacing}{\singlespacing}
\makeatother

\normalspacing

\makeatletter%
\def\nottoobig#1{{\hbox{$\left#1\vcenter to1.111\ht\strutbox{}\right.\n@space$}}}
\makeatother%

\makeatletter 

\newcommand{\wh}[1]{ \widehat{#1} }

\newlength{\filength}
\newsavebox{\gcbox}
\sbox{\gcbox}{\framebox[\filength]{\rule{0ex}{2ex}}}

\newlength{\leftjustindent}
\newlength{\@leftjustindent}
\def\leftjust{\let\\\@leftjustcr\let\end\@endleftjust
  \addtolength{\@leftjustindent}{\leftjustindent}
  \vcenter\bgroup
  \halign\bgroup
    \hbox to\displaywidth{
      \rule{\@leftjustindent}{0ex}$\displaystyle##$\hfill
      }\crcr
}
\def\endleftjust{\crcr\egroup\egroup\endgroup}
\def\@endleftjust#1{\crcr\egroup\egroup\@checkend{#1}\endgroup}
\def\@leftjustcr{\crcr}

\newtheorem{theorem}{Theorem}[section]
\newtheorem{corollary}[theorem]{Corollary}
\newcommand{\qedblob}{\mbox{\rule[-1.5pt]{5pt}{10.5pt}}}
\renewcommand\qedblob{\ding{113}}
\def\literalqed{{\ \nolinebreak\hfill\mbox{\qedblob\quad}}}

\def\qed{\literalqed}

\newtheorem{proposition}[theorem]{Proposition}
\makeatother

\newtheorem{definition}[theorem]{Definition}

\newcommand{\sat}{{\rm SAT}}

\newcommand{\p}{{\rm P}}

\newcommand{\np}{{\rm NP}}

\def\pair#1{{{\langle\!\!~#1~\!\!\rangle}}}

\newcommand{\calc}{\mbox{$\cal C$}}

\newcommand{\condition}{\,\mid\:}
\def\land{{\; \wedge \;}}
\newcommand{\fp}{{\rm FP}}

\newcommand{\ddash}{\penalty10000{\mathchoice{\hbox{\normalsize-}}{\hbox{\normalsize-}}{\hbox{\scriptsize-}}{\hbox{\tiny-}}}}
\newcommand{\dash}{\ifmmode\protect\ddash\else-\fi}

\newcommand{\cminus}{\,-\,}

\newcommand{\etime}{{\mbox{\rm E}}}

\title{$\p$-Immune Sets with Holes Lack Self-Reducibility Properties}

\author{%
Lane A. Hemaspaandra\thanks{%
\protect\singlespacing
{\tt lane@cs.rochester.edu}.  Supported in part by grants
NSF-INT-9815095/\protect\linebreak[0]DAAD-315-PPP-g\"u-ab and 
NSF-CCR-9322513.  
}
\\
Department of Computer Science\\
University of Rochester\\
Rochester, NY 14627-0226, USA
\and 
Harald Hempel\thanks{%
\protect\singlespacing
{\tt hempel@informatik.uni-jena.de}.
Supported in part 
by grant
NSF-INT-9815095/\protect\linebreak[0]DAAD-315-PPP-g\"u-ab.
Work done while visiting the University of Rochester.
}
\\Institut f\"ur Informatik\\
Friedrich-Schiller-Universit\"at Jena\\
D-07743 Jena, Germany
}

\date{%
February 23, 2001}

\begin{document}

\sloppy

\bibliographystyle{alpha}

\maketitle

\begin{abstract}
No $\p$-immune set having exponential gaps is
positive-Turing self-reducible.
\end{abstract}

\section{Introduction}\label{s:intro}

A set is $\p$-immune if it is infinite yet has no infinite P
subsets (\cite{ber:c:complete-sets}, 
see also~\cite{ben-gil:j:prob1,ko-moo:j:approx}). 
That is, each $\p$-immune set is so hard that any
polynomial-time machine that avoids accepting elements of its
complement can in fact recognize only a finite number of its elements.
Informally put, P sets cannot well-approximate it from the inside.
$\p$-immunity has been extensively studied in complexity theory.

We are interested in the following issue: Does $\p$-immunity have any
repercussions regarding self-reducibility properties? In particular,
does $\p$-immunity ever preclude certain self-reducibilities? There are
two papers in the literature that prove results in this direction.

The paper ``Strong Self-Reducibility Precludes Strong
Immunity''~\cite{hem-zim:j:balanced} proves that the complexity
class NT is not $\p$-balanced-immune. NT, the near-testable
set~\cite{gol-hem-jos-you:j:nt}, 
is a class made up exactly of sets having a very
specific 1-query-fixed-truth-table self-reducibility property. That
paper thus realizes the on-target intuition, due to Eric Allender (see
the acknowledgments section of~\cite{hem-zim:j:balanced}), that ``the
restrictive self-reducibility structure of NT [should] constrain one's
ability to achieve strong separations from P''~\cite{hem-zim:j:balanced}.

The other paper exploring the extent to which $\p$-immunity might
conflict with self-reducibility properties is a 1990 paper by K\"amper
that proves that $\p$-immune sets having 
\emph{double-exponentially} large holes
can never be \emph{disjunctive-Turing} 
self-reducible~\cite{kae:j:self-reducibility}. 
Our work was motivated by K\"amper's paper and by the 
desire to see in what further ways $\p$-immunity may preclude
self-reducibility.

Our main result is that $\p$-immune sets having \emph{exponentially} 
large holes can
never be \emph{positive-Turing} self-reducible (or even locally
left-positive-Turing \emph{word-decreasing}-self-reducible). We also prove
other related results.

\section{Preliminaries}\label{s:prelim}

We assume the reader to be familiar with the basic concepts of complexity 
theory~\cite{pap:b:complexity, bal-dia-gab:b:sctI-2nd-ed}.
Throughout the paper all logarithms are base 2. 
The following reduction types will be used in this paper.

\begin{definition}\label{d:reductions}
Let $A$ and $B$ be sets and $M$ be a Turing machine.
\begin{enumerate}
\item (see~\cite{lad-lyn-sel:j:com})\quad
We say that $A \leq_{Turing}^p B$ via $M$ 
(``$A$ Turing reduces to $B$ via $M$'')
if 
$M$ is a deterministic polynomial-time Turing machine and $A=L(M^B)$.
\item \cite{sel:j:reductions-pselective}\quad 
We say that $A \leq_{positive\hbox{-}Turing}^p B$ via $M$ 
(``$A$ positive-Turing reduces to $B$ via $M$'')
if 
$M$ is a deterministic polynomial-time Turing machine,
$A=L(M^B)$, and for all sets $C$ and $D$ such that $C \subseteq D$ 
it holds that $L(M^C) \subseteq L(M^D)$.   
\item (see~\cite{lad-lyn-sel:j:com})\quad 
We say that $A \leq_{disjunctive\hbox{-}Turing}^p B$ via $M$ 
(``$A$ disjunctive-Turing reduces to $B$ via $M$'')
if $M$ is a deterministic polynomial-time Turing machine,
$A=L(M^B)$, and for all $x$, $x \in A$ if and only if 
$M^B(x)$ generates at least one query that is a member of $B$.
\item \cite{hem-jai:j:pos}\quad 
We say that $A \leq_{lpos\hbox{-}Turing}^p B$ via $M$ 
(``$A$ locally left-positive-Turing reduces to $B$ via $M$'')
if $M$ is a deterministic polynomial-time Turing machine,
$A=L(M^B)$, and for all sets $C$,  
$L(M^{B\cminus C})\subseteq L(M^B)$.
\end{enumerate}
\end{definition}

Self-reducibility is a central notion in complexity 
theory (see~\cite{jos-you:b:internal-structure}). 
It appeared in concept first in Schnorr~\cite{sch:c:self-reducible},
and was formalized and extended by Meyer and 
Paterson~\cite{mey-pat:t:int},
Balc\'azar~\cite{bal:j:self}, and others.

\begin{definition}\label{d:self-reductions}
Let $A$ and $B$ be sets.
\begin{enumerate}
\item 
For any $r$ for which ``$A \leq_{r}^p B$ via $M$'' 
has been defined, 
$A \leq_{r}^p B$ is defined as meaning there is a deterministic
polynomial-time machine $M$ such that $A \leq_{r}^p B$ via $M$.
\item 
(see~\cite{bal:j:self, bal-dia-gab:b:sctI-2nd-ed})\quad
For any $r \neq disjunctive\mbox{-}Turing$ for which 
``$A \leq_{r}^p B$ via $M$'' 
has been defined, 
$A$ is said to be $r$~self-reducible 
if there is a deterministic polynomial-time machine $M$ such that
\begin{enumerate}
\item $A \leq_{r}^p A$ via $M$, and 
\item on each input $x$, $M^A(x)$ queries only strings of length
strictly less than $|x|$.
\end{enumerate}
\item 
(see~\cite{bal:j:self})\quad
For any $r \neq disjunctive\mbox{-}Turing$ 
for which ``$A \leq_{r}^p B$ via $M$'' 
has been defined, 
$A$ is said to be $r$~word-decreasing-self-reducible if 
there is a deterministic polynomial-time machine $M$ such that
\begin{enumerate}
\item $A \leq_{r}^p A$ via $M$, and 
\item on each input $x$, $M^A(x)$ queries only strings that
are lexicographically strictly less than $x$.
\end{enumerate}
\end{enumerate}
\end{definition}

Under the above definition,
if we had not in it  
explicitly excluded
the case of disjunctive-Turing reductions, 
only the empty set would be 
disjunctive-Turing self-reducible and only the empty set would be 
disjunctive-Turing word-decreasing self-reducible.  
The reason is that
there is no way to get a ``first'' string into the set.
Many textbooks are a bit careless on this point.
However, careful definitions, such as that of 
Ambos-Spies and K\"amper~\cite{amb-kae:c:self-reducible-degrees} of 
disjunctive-Turing self-reducibility, avoid this problem.  
(The same issue of course exists regarding
disjunctive-truth-table self-reducibility, conjunctive-truth-table
self-reducibility, and conjunctive-Turing self-reducibility, and 
is handled analogously.)

\begin{definition}
\label{d:disjunctive-self}
Let $A$ and $B$ be sets and $M$ be a Turing machine.
\begin{enumerate}
\item We say that $A \leq_{disjunctive'\hbox{-}Turing}^p B$ via $M$
(``$A$ disjunctive${\,}^\prime$-Turing reduces to $B$ via $M$'')
if $M$ is a deterministic polynomial-time Turing machine,
$A = L(M^B)$, and 
$M$ has the following acceptance behavior:  
on each 
input it accepts exactly if either (a)~it halted in an 
accepting state without asking any queries, or (b)~it asked
at least one query and at least one query it asked received
the answer ``yes.''

\item(essentially~\cite{amb-kae:c:self-reducible-degrees})\quad
A set $B$ is said to be disjunctive-Turing self-reducible %
if there is a deterministic
polynomial-time machine $M$ such that
\begin{enumerate}
\item $B \leq_{disjunctive'\hbox{-}Turing}^p B$ via $M$, and  %
\item  on each input $x$, $M^B(x)$ queries only strings of length
strictly less than $|x|$.
\end{enumerate}
 
\item 
A set $B$ is said to be disjunctive-Turing 
word-decreasing-self-reducible if there is a deterministic
polynomial-time machine $M$ such that
\begin{enumerate}
\item $B \leq_{disjunctive'\hbox{-}Turing}^p B$ via $M$, and
\item on each input $x$, $M^B(x)$ queries only strings that
are lexicographically strictly less than $x$.
\end{enumerate}
\end{enumerate}
\end{definition}

Of course, for each $r$, every $r$~self-reducible set is
$r$~word-decreasing-self-reducible.  (For explicitness, in some of our
theorems that apply to both we will mention both in the theorem
statements.)

If one wishes to define conjunctive-Turing self-reducibility and 
conjunctive-Turing word-decreasing self-reducibility one,
for reasons analogous to those outlined above,   
has to make the same type of special case as is done 
for disjunctive-Turing 
self-reducibility  in Definition~\ref{d:disjunctive-self}.

It is well-known that $A \leq_{disjunctive\hbox{-}Turing}^p B$ implies $A
\leq_{positive\hbox{-}Turing}^p B$ which in turn implies $A
\leq_{lpos\hbox{-}Turing}^p B$.  
And there exist sets $A',B',A'',B''$ such that 
$A'\leq_{lpos\hbox{-}Turing}^p B'$ yet 
$A'\not\leq_{positive\hbox{-}Turing}^p B'$~\cite{hem-jai:j:pos},
and  
$A''\leq_{positive\hbox{-}Turing}^p B''$ yet 
$A''\not\leq_{disjunctive\hbox{-}Turing}^p B''$~\cite{lad-lyn-sel:j:com}.
That is, 
$\leq_{lpos\hbox{-}Turing}^p$ is a more broadly applicable reduction than
$\leq_{positive\hbox{-}Turing}^p$, which in turn is a more broadly applicable 
reduction than $\leq_{disjunctive\hbox{-}Turing}^p$.

Self-reducible sets have been intensively studied.  
It is well-known that all disjunctive-Turing self-reducible sets are in $\np$
and also in $\etime=\bigcup_{c>0} {\rm DTIME}(2^{cn})$.  Balc\'azar showed
that in fact every Turing word-decreasing-self-reducible set is in 
$\etime$.

\begin{theorem} 
\label{t:selfred-e}
\cite{bal:j:self}\quad 
Every Turing word-decreasing-self-reducible set is in $\etime$.
\end{theorem}

Immunity is a concept developed to study the degree of separation that can
be achieved between classes (see~\cite{rog:b:rft}).
In particular, $\p$-immunity~\cite{ber:c:complete-sets}
(see also~\cite{ben-gil:j:prob1,ko-moo:j:approx})
is a well-studied concept.

\begin{definition}
(see~\cite{rog:b:rft})\quad 
Let $\calc$ be any class.
A set $B$ is called $\calc$-immune if $B$ is infinite yet
no infinite subset of $B$ belongs to $\calc$.
\end{definition}

\begin{definition}
A set $A$ has exponential-size gaps (E-gaps) if the following holds:
$$(\exists c>0)(\forall n\in \mathbb{N})(\exists m>n)
[\{z \in A\condition m < |z| \leq 2^{cm}\}= \emptyset].$$
\end{definition}

A set $A$ has double-exponential-size gaps  if 
$(\forall n\in \mathbb{N})(\exists m \geq n)
[\{z \in A\condition m< |z| \leq 2^{2^m}\}= \emptyset]$.
It has been shown by K\"amper~\cite{kae:j:self-reducibility} that 
no $\p$-immune set $A$ having double-exponential-size gaps
can 
be disjunctive-Turing self-reducible.
K\"amper proves his result for the model, different than
that of this paper, in which self-reducibility is defined with respect
to all polynomially well-founded orders.
In Section~\ref{s:results} we show that, in fact, no $\p$-immune set
having exponential-size gaps is positive-Turing self-reducible (or
even locally left-positive-Turing word-decreasing-self-reducible).
In Section~\ref{s:hardness}, we study related issues such 
as consequences for SAT that would follow from NP-hardness for
sets having exponential-sized gaps.

\section{Immunity with Holes versus Self-Reducibility}
\label{s:results}

We now state our main theorem.

\begin{theorem}\label{t:main}
No $\p$-immune set having E-gaps is 
locally left-positive-Turing word-decreasing-self-reducible.
\end{theorem}

\begin{corollary}
\begin{enumerate}
\item No $\p$-immune set having E-gaps is positive-Turing  
self-reducible or positive-Turing word-decreasing-self-reducible. 
\item No $\p$-immune set having E-gaps is disjunctive-Turing  
self-reducible or disjunctive-Turing word-decreasing-self-reducible.
\item No $\p$-immune set having E-gaps
is locally left-positive-Turing self-reducible.
\end{enumerate}
\end{corollary}

{\bf Proof of Theorem~\ref{t:main}:}\quad
Let $A$ be a locally left-positive-Turing word-decreasing-self-reducible set 
having $E$-gaps. 
If $A$ is finite, it is trivially not $\p$-immune.
So suppose that $A$ is infinite. 
It suffices to show that $A$ has an infinite subset in $\p$.

Let $c>0$ be a constant such that $(\forall n\in \mathbb{N})(\exists m>n)
[\{z \in A\condition m< |z| \leq 2^{cm}\}= \emptyset]$. 
Let $M$ be a deterministic polynomial-time Turing machine 
witnessing the locally left-positive-Turing word-decreasing-self-reducibility 
of $A$, in other words, 
\begin{enumerate}
\item $L(M^A)=A$,
\item for all sets $C$,  $L(M^{A\cminus C})\subseteq L(M^A)$, and 
\item on each input $x$, $M^A(x)$ queries only strings that
are lexicographically strictly less than $x$.
\end{enumerate}
By Theorem~\ref{t:selfred-e}, there exist a constant $d>0$ and a 
deterministic $2^{dn}$-time-bounded Turing machine $M_e$ such that 
$L(M_e)=A$.

Consider the following deterministic Turing machine $M'$:
\begin{enumerate}
\item On input $x$ simulate the action of $M(x)$ while 
answering the queries generated during that simulation  as follows:
\begin{enumerate}
\item Every query $q$ with $|q|\leq \frac{\log |x|}{c}$ is answered 
according to the outcome of $M_e(q)$, 
i.e., if $|q|\leq \frac{\log |x|}{c}$ then $M_e(q)$ is simulated and the 
query $q$ generated by $M(x)$ is answered ``yes'' if 
$M_e(q)$ accepts and is answered ``no'' otherwise.
\item Every query $q$ with $\frac{\log |x|}{c} <|q|$ is answered ``no.''
\end{enumerate}
\item Accept if and only if the simulation of $M(x)$, answering the 
queries 
(generated during the simulation of $M(x)$) 
as described above, accepts. 
\end{enumerate}

It is not hard to see that $M'(x)$ runs in time polynomial in $|x|$.
Let $B=L(M')$. It follows that $B \in \p$. 

\begin{description}
\item[Claim 1] $B\subseteq A$.

Let $x$ be a string such that $x \in B$, in other words, $x \in L(M')$.
Since $L(M^A)=A$ and $L(M_e)=A$, 
$M'(x)$ gets the right answer to each query $q$ that $M(x)$ generates 
that satisfies $|q| \leq \frac{\log |x|}{c}$. 
Since all other queries are answered ``no,'' the 
outcome of $M'(x)$ is identical to the outcome of 
$M^{\left(A^{\leq \frac{\log |x|}{c}}\right)}(x)$.
Thus $x \in L\left(M^{\left(A^{\leq \frac{\log |x|}{c}}\right)}\right)$.
But note that due to the properties of $M$,
$L\left(M^{\left(A^{\leq \frac{\log |x|}{c}}\right)}\right)\subseteq L(M^A)$. 
Hence $x \in A$.

\item[Claim 2] $B$ is infinite.\\
Let $m_0<m_1<m_2<\dots$ be an infinite sequence of natural numbers 
such that for all $i\geq 0$, both 
$\{z \in A\condition m_i< |z| \leq 2^{cm_i}\}= \emptyset$ 
and 
$\{z \in A\condition 2^{cm_i} < |z| \leq m_{i+1}\}\not= \emptyset$ hold.
Such a sequence exists since $A$ is infinite and has $E$-gaps.
For all $i\geq 0$, define 
$z_i=\min\{z \in A\condition 2^{cm_i} < |z| \leq m_{i+1}\}$,
where we minimize with respect to the lexicographical order on strings. 
Note that for all $i\geq 0$, due to $\frac{\log|z_i|}{c} >m_i$ 
all strings $y$, $y<_{lex} z_i$, of length $|y|>\frac{\log|z_i|}{c}$ 
are not in $A$ since they fall into the gap that extends at least down 
to the length $m_i+1$ and that stretches at least up to just before $z_i$.
Informally, $z_i$ is the (lexicographically) first string in $A$ beyond one of 
$A$'s exponential-size gaps.

It is clear that $\{z_i\condition i\geq 0\}$ is an infinite set. 
In order to show that $B$ is infinite it certainly suffices to show that 
$\{z_i\condition i\geq 0\}\subseteq B$.
But this follows from the fact that by construction for all $i\geq 0$  
during the run of 
$M'(z_i)$ we correctly simulate the work of $M^A(z_i)$ since the answers 
to all queries $q$, $|q| \leq \frac{\log |z_i|}{c}$, generated by $M^A(z_i)$ 
are correctly found with the help of $M_e(q)$ and the answers 
to all queries $q$, $|q| > \frac{\log |z_i|}{c}$, generated by $M^A(z_i)$ 
are truly ``no'' since all those queries fall into the gap above the 
length $m_i$.

This completes the proof of Claim~2 and the proof of the theorem.~\qed
\end{description}

\section{Emptiness Testing and 
\boldmath$\np$-Hardness\unboldmath}\label{s:hardness}

In order to extend Theorem~\ref{t:main} to arbitrary Turing  
self-reducible sets---as opposed to requiring positivity properties
for the self-reducibility---it 
appears crucial to have some 
knowledge of where the set has its holes.

\begin{definition}
A set $B$ is said to be emptiness-testable if and only if 
$$\{1^i \condition B^{=i}=\emptyset\} \in \p.$$
\end{definition}

\begin{proposition}
\label{p:emptyinterval}
A set $B$ is emptiness-testable if and only if 
$$EmptyInterval_B = \{0^i1^j\condition i\leq j
\land\{z\in B\condition i\leq |z|\leq j\}=\emptyset\}\in \p.$$
\end{proposition}

The proof of the proposition is immediate.

\begin{theorem}\label{t:main-empty}
No $\p$-immune emptiness-testable set having E-gaps is Turing 
word-decreasing-self-reducible.
\end{theorem}

\begin{proof}
The proof is similar to the proof of Theorem~\ref{t:main}. 
Let $A$ be an infinite emptiness-testable Turing 
word-decreasing-self-reducible set having $E$-gaps.
If $A$ is finite then $A$ is not $\p$-immune, 
so we henceforth consider only the case that $A$ is infinite.
We will show that $A$ has an infinite subset in $\p$.

Let $c>0$ be a constant such that $(\forall n\in \mathbb{N})(\exists m>n)
[\{z \in A\condition m< |z| \leq 2^{cm}\}= \emptyset]$. 
Let $M$ be a deterministic polynomial-time Turing machine 
witnessing the Turing word-decreasing-self-reducibility of $A$.
By Theorem~\ref{t:selfred-e}, there exist a constant $d>0$ and a 
deterministic $2^{dn}$-time-bounded Turing machine $M_e$ such that 
$L(M_e)=A$.
Let $T_e=\{1^i\condition A^{=i}=\emptyset\}$. 
By assumption $T_e\in \p$.

Consider the following deterministic Turing machine $M'$:
\begin{enumerate}
\item On input $x$ simulate the work of $M(x)$ while 
answering the generated queries as follows: 
(Initialize $u=1$. 
The variable $u$ will work as a flag to indicate whether the answers to 
certain oracle queries are correct.)

\begin{enumerate}
\item 
Every query $q$ with $|q|\leq \frac{\log |x|}{c}$ is answered 
according to the outcome of $M_e(q)$, 
i.e., simulate $M_e(q)$ and answer ``yes'' to the query $q$ 
if $M_e(q)$ accepts and answer ``no'' otherwise.
\item 
Every query $q$ with $\frac{\log |x|}{c} <|q|$ is answered ``no.''
If $1^{|q|} \not\in T_e$, 
that is if $A^{=|q|}\not=\emptyset$, 
set $u=u+1$, otherwise leave $u$ unchanged.
(Informally put, we change the value of $u$ if we answered ``no'' to a query 
that is of a length at which $A$ is \emph{not} empty.)
\end{enumerate}
\item Accept if and only if both the simulation of $M(x)$ while answering the 
queries as described above accepts and $u=1$. 
\end{enumerate}

From here on the proof proceeds in analogy to the proof of 
Theorem~\ref{t:main}.~\qed
\end{proof}

It is not hard to see that the information about the emptiness of a set $A$ 
can also be present in form of one bit of advice \emph{per length}. 
Thus we have the following corollary. 
(Note, $\p/1 =\{L\condition 
(\exists B\in \p)(\exists f:1^{\ast}\rightarrow\{0,1\})(\forall x)
[x\in L \iff \pair{x,f(1^{|x|})} \in B]\}$, see~\cite{kar-lip:c:nonuniform}.)

\begin{corollary}
No $\p/1$-immune set having E-gaps can be Turing 
word-decreasing-self-reducible or Turing self-reducible.
\end{corollary}

We mention in passing that the proof of Theorem~\ref{t:main-empty} 
shows something slightly stronger than claimed in the statement of 
Theorem~\ref{t:main-empty}, namely, 
that any 
self-reducible set having E-gaps can be $\p$-immune only if its gaps are 
``hard to find,'' in other words, no $\fp$ function should be able to 
recognize an infinite number of its gaps.

Can emptiness-testable sets having E-gaps be $\np$-hard? 
If this would be the case than it would follow from Theorem~\ref{t:main-empty}
that such an $\np$-hard set could not be $\p$-immune. 
A few definitions will be helpful in studying the above question.
By $\np$-hard we always mean 
$\leq^p_m$-hard for $\np$.

\begin{definition}\label{d:easyband}
\begin{enumerate}
\item
Let $\calc$ be a complexity class.
A set $D$ has $\calc$-easiness bands if, for 
every 
$\ell > 1 $, there exists 
a set $B\in \calc$  such that, for infinitely many $n \in
\mathbb{N}$,
$$
(D \Delta B)\cap
\{z \condition n\leq |z|\leq n^{\ell}\}=\emptyset,$$
where $\Delta$ denotes the symmetric difference of sets, i.e.,
$D \Delta B=(D\cminus B)\cup(B \cminus D)$.
\item \label{d:easy-two}
Let $\calc$ be any complexity class.
A set $D$ is said to have obvious $\calc$-easiness bands
if, for every  $\ell > 1$, there exist a set $B\in \calc$ and
an infinite
tally set $T \in {\cal C}$
such that, for all $1^n \in T$,
$$
(D \Delta B)\cap
\{z \in \condition n\leq |z|\leq n^{\ell}\}=\emptyset.$$
\end{enumerate}
\end{definition}

If we wanted to apply part~\ref{d:easy-two} of
Definition~\ref{d:easyband} to classes much less nicely
behaved than $\p$, we would want to replace the
tally-set-$T \in {\cal C}$ condition with 
a requirement that the tally set be
${\cal C}$-printable with respect to some natural 
printability notion corresponding 
to~$\cal C$~\cite{har-yes:j:computation}.  However, in
this paper we will use the definition only as applied to~$\p$.

\begin{theorem}
\begin{enumerate}
\item If any set in $\etime$ having E-gaps is $\np$-hard, 
then $\sat$ (and indeed
all positive-Turing word-decreasing-self-reducible $\np$ sets)
has $\p$-easiness bands.
\item If any emptiness-testable set in $\etime$ having E-gaps is $\np$-hard, 
then $\sat$ 
(and indeed all positive-Turing word-decreasing-self-reducible $\np$ sets)
has obvious $\p$-easiness bands.
\end{enumerate}
\end{theorem}

\begin{proof}
Regarding part~1 of the theorem, 
let $A$ be a set in $\etime$ having E-gaps. 
Let $c>0$ be a constant such that for infinitely many $m\in \mathbb{N}$ it 
holds that $\{z \in A\condition m< |z| \leq 2^{cm}\}= \emptyset$.
Since $A\in \etime$ there exist a constant $d>0$ and a 
$2^{dn}$-time-bounded Turing machine $M_e$ such that $L(M_e)=A$.
Suppose that $A$ is  $\np$-hard. 
Let $f$ be a polynomial-time computable such reduction, 
i.e., for all $x$, $x\in \sat \iff f(x)\in A$.
Since $\sat$ is positive-Turing self-reducible 
(even disjunctive-Turing self-reducible) 
there exists a deterministic polynomial-time machine $M$ such that 
\begin{enumerate}
\item $\sat=L(M^\sat)$, 
\item on each input $x$, $M^A(x)$ queries only strings of length
strictly less than $|x|$, and 
\item for all $C$ and $D$ such that 
$C \subseteq D$ it holds that $L(M^C) \subseteq L(M^D)$. 
\end{enumerate}

We will show that, for every $\ell >1$, there exists a $\p$ set $B$
such that, for infinitely many $n \in \mathbb{N}$,
$$
(\sat \Delta B)\cap 
\{z \condition n\leq |z|\leq n^{\ell}\}=\emptyset.
$$
We will do so by showing that, for every $k >1$, there exist a  $\p$ set 
$B'\subseteq \sat$ and an infinite tally set $C$ such that 
for all $1^n \in C$,
$$
\sat \cap 
\{z \condition n^{\frac{1}{k}}\leq |z|\leq n\}\subseteq B'.
$$
Though the format here is $n^{\frac{1}{k}}$ versus $n^1$ rather 
than $n^1$ versus $n^{\ell}$,
it is not hard to see that this suffices. 
Let $p$ be polynomial such that for all $x$ and for all 
$n \in \mathbb{N}$, $|f(x)|\leq p(|x|)$ and $p(n) < p(n+1)$. 

Let $k>1$. 
Consider the following deterministic Turing machine $M'$:

\begin{enumerate}
\item 
On input $x$, $|x|=n$, 
simulate $M(x)$ and  each time $M(x)$ asks a query $q$ to 
$\sat$ compute $f(q)$ and answer the query ``$q \in \sat?$'' as follows:
\begin{enumerate}
\item 
If  
$|f(q)|\leq \frac{\log(n^kp(n^k))}{c}$ 
then answer ``yes'' if and 
only if  $M_e(f(q))$ accepts and ``no'' otherwise.
\item If 
$\frac{\log(n^kp(n^k))}{c} <|f(q)|$ 
then answer ``no.''
\end{enumerate}
\item Accept if and only if the simulation of $M(x)$, answering the 
queries as described above, accepts. 
\end{enumerate}

It is not hard to see that the above machine $M'$ runs in time polynomial in 
$n$.
Let $B'=L(M')$.
Since $M$ is globally positive and the above machine 
answers queries by exploiting the 
many-one reduction from $\sat$ to $A$ or by answering ``no,'' 
it follows that $B' \subseteq \sat$.

Define $C=\{1^n\condition (\exists m\in \mathbb{N})
[\{z \in A\condition m< |z| \leq 2^{cm}\}= \emptyset \land 
p(n) <2^{cm} \leq np(n)]\}$. 
Note that $C$ is infinite. 
To see this let $\wh{n}$ be such that, for all $n\geq \wh{n}$,
$p(n+1)<np(n)$. Such an $\wh{n}$ clearly exists, since $p$ is 
a monotonic polynomial of degree greater than zero.
Now let $m$ be 
any natural number such that 
$p(\wh{n}) <  2^{cm}$ and 
$\{z \in A\condition m< |z| \leq 2^{cm}\}= \emptyset$.  
Define $n_m =\max\{n'\condition p(n') <2^{cm}\}$. 
Note $n_m \geq \wh{n}$ and
$2^{cm} \leq p(n_m+1) < n_mp(n_m)$. 
It follows that $n_m \in C$. 
Since there are infinitely many $m$ satisfying both, $p(\wh{n}) < 2^{cm}$ 
and $\{z \in A\condition m< |z| \leq 2^{cm}\}= \emptyset$,
 it follows that $C$ 
is an infinite set.

We are now prepared to show that for all $1^n \in C$, 
$$\sat \cap 
\{z \condition n^{\frac{1}{k}}\leq |z|\leq n\}\subseteq B'.$$

Let $1^n \in C$.  In light of the definition 
of $C$, 
there exists some $m\in \mathbb{N}$ such that 
$\{z \in A\condition m< |z| \leq 2^{cm}\}= \emptyset$ and 
$p(n) <2^{cm} \leq np(n)$.   Choose such an $m$ (which implicitly
is $m_n$).
Note that $
2^{cm}\leq np(n)$ implies $m\leq \frac{\log(np(n))}{c}$. 
Hence any string $y$ satisfying $\frac{\log (np(n))}{c} <|y|\leq p(n)$ cannot 
be in $A$. 

Let $z$ be such that 
$n^{\frac{1}{k}}\leq |z|\leq n$ and suppose that $z \in \sat$. 
Note that $n\leq |z|^k$. 
So, since $\log$ and $p$ are monotonic, 
$m\leq \frac{\log(|z|^kp(|z|^k))}{c}$, 
and of course $p(|z|)\leq p(n) <2^{cm}\leq np(n) \leq |z|^kp(|z|^k)$.
This implies that any string $y$ satisfying 
$\frac{\log(|z|^kp(|z|^k))}{c} <|y|\leq p(|z|)$ cannot be in $A$. 

Now consider the action of $M'(z)$. $M'(z)$ essentially simulates the work of 
$M(z)$. 
Note that for all queries $q$
generated by $M(z)$, $|q|\leq |z|$ and hence 
$|f(q)|\leq p(|z|)$.  
Furthermore, any
query $q$ with $|f(q)|\leq \frac{\log(|z|^kp(|z|^k))}{c}$ is correctly
answered during the simulation of $M(z)$ in our algorithm since $L(M_e)=A$.  
On the other hand, for all queries $q$ with 
$|f(q)| >\frac{\log(|z|^kp(|z|^k))}{c}$ 
(recall that those queries are answered ``no'' by $M'(z)$ 
during the simulation of $M(z)$) 
$f(q)$ is in the gap associated with $m$ 
(i.e., the gap that extends at least down to the length $m+1$ and stretches at least up to the length $2^{cm}$), 
in other words, 
$f(q)\not\in A$ and consequently $q\not\in\sat$.  
This shows that during the run of $M'(z)$ all queries generated in the 
simulation of $M(z)$ are answered correctly and hence 
$z \in \sat$ implies $z\in B'$.

So we showed that, under the assumption of part~1 of the theorem,
$\sat$ has $\p$-easiness bands. 
The same proof works for any positive-Turing self-reducible $\np$ set, 
or indeed, 
with the obvious minor change in the proof, 
for any positive-Turing word-decreasing-self-reducible $\np$ set.
This completes the proof of part~1.

Regarding the proof of part~2 we note that if $A$ is emptiness-testable, 
then the above-defined set $C$ is in $\p$. 
This can be seen easily in light of the definition of $C$, 
using also Proposition~\ref{p:emptyinterval}.
Though the set $C$ of this proof marks upper ends of bands
in contrast with part~\ref{d:easy-two}
of Definition~\ref{d:easyband} which requires the marking of 
the lower ends, it is not hard to see that this suffices,
though due to rounding issues one has to be slightly careful.
In particular, if we wish to prove bands of the form
$n$-to-$n^{\ell}$, we use the above proof for the value 
$k = \ell + 1$ to get bands of the form 
$n^{1/(\ell + 1)}$-to-$n$ and to get an upper-edge-marking set $C \in \p$.
The set $C' = \{ 1^{\lceil j^{1/(\ell + 1) }\rceil} 
\condition 1^j \in C \mbox{ and } 
{\lceil j^{1/(\ell + 1)}\rceil}^{\ell} \leq j\}$ 
will 
also be in $\p$, will be infinite, and will serve as the 
desired lower-edge-of-band marking tally set in the sense of 
part~\ref{d:easy-two}
of Definition~\ref{d:easyband}~\qed 
\end{proof}

\bibliography{gry}

\begin{thebibliography}{GHJY91}

\bibitem[AK88]{amb-kae:c:self-reducible-degrees}
K.~{Ambos-Spies} and J.~K{\"{a}}mper.
\newblock On disjunctive self-reducibility.
\newblock In {\em Proceedings of the 2nd Workshop on Computer Science Logic},
  pages 1--13. Springer-Verlag {\it Lecture Notes in Computer Science \#385},
  October 1988.

\bibitem[Bal90]{bal:j:self}
J.~Balc{\'{a}}zar.
\newblock Self-reducibility.
\newblock {\em Journal of Computer and System Sciences}, 41(3):367--388, 1990.

\bibitem[BDG95]{bal-dia-gab:b:sctI-2nd-ed}
J.~Balc{\'{a}}zar, J.~D{\'{\i}}az, and J.~Gabarr{\'{o}}.
\newblock {\em Structural Complexity I}.
\newblock EATCS Texts in Theoretical Computer Science. Springer-Verlag, 2nd
  edition, 1995.

\bibitem[Ber76]{ber:c:complete-sets}
L.~Berman.
\newblock On the structure of complete sets.
\newblock In {\em Proceedings of the 17th IEEE Symposium on Foundations of
  Computer Science}, pages 76--80, 1976.

\bibitem[BG81]{ben-gil:j:prob1}
C.~Bennett and J.~Gill.
\newblock Relative to a random oracle {${A}$}, ${{\rm P}^A} \not = {{\rm NP}^A}
  \not = {{\rm coNP}^A}$ with probability 1.
\newblock {\em SIAM Journal on Computing}, 10:96--113, 1981.

\bibitem[GHJY91]{gol-hem-jos-you:j:nt}
J.~Goldsmith, L.~Hemachandra, D.~Joseph, and P.~Young.
\newblock Near-testable sets.
\newblock {\em SIAM Journal on Computing}, 20(3):506--523, 1991.

\bibitem[HJ91]{hem-jai:j:pos}
L.~Hemachandra and S.~Jain.
\newblock On the limitations of locally robust positive reductions.
\newblock {\em International Journal of Foundations of Computer Science},
  2(3):237--255, 1991.

\bibitem[HY84]{har-yes:j:computation}
J.~Hartmanis and Y.~Yesha.
\newblock Computation times of {NP} sets of different densities.
\newblock {\em Theoretical Computer Science}, 34:17--32, 1984.

\bibitem[HZ96]{hem-zim:j:balanced}
L.~Hemaspaandra and M.~Zimand.
\newblock Strong self-reducibility precludes strong immunity.
\newblock {\em Mathematical Systems Theory}, 29(5):535--548, 1996.

\bibitem[JY90]{jos-you:b:internal-structure}
D.~Joseph and P.~Young.
\newblock Self-reducibility: Effects of internal structure on computational
  complexity.
\newblock In A.~Selman, editor, {\em Complexity Theory Retrospective}, pages
  82--107. Springer-Verlag, 1990.

\bibitem[K{\"{a}}m90]{kae:j:self-reducibility}
J.~K{\"{a}}mper.
\newblock A result relating disjunctive self-reducibility to {P}-immunity.
\newblock {\em Information Processing Letters}, 33(5):239--242, 1990.

\bibitem[KL80]{kar-lip:c:nonuniform}
R.~Karp and R.~Lipton.
\newblock Some connections between nonuniform and uniform complexity classes.
\newblock In {\em Proceedings of the 12th ACM Symposium on Theory of
  Computing}, pages 302--309. ACM Press, April 1980.
\newblock An extended version has also appeared as: Turing machines that take
  advice, {\em L'Enseignement Math\'{e}matique}, 2nd series, 28, 1982,
  pages~191--209.

\bibitem[KM81]{ko-moo:j:approx}
K.~Ko and D.~Moore.
\newblock Completeness, approximation, and density.
\newblock {\em SIAM Journal on Computing}, 10(4):787--796, 1981.

\bibitem[LLS75]{lad-lyn-sel:j:com}
R.~Ladner, N.~Lynch, and A.~Selman.
\newblock A comparison of polynomial time reducibilities.
\newblock {\em Theoretical Computer Science}, 1(2):103--124, 1975.

\bibitem[MP79]{mey-pat:t:int}
A.~Meyer and M.~Paterson.
\newblock With what frequency are apparently intractable problems difficult?
\newblock Technical Report MIT/LCS/TM-126, Laboratory for Computer Science,
  MIT, Cambridge, MA, 1979.

\bibitem[Pap94]{pap:b:complexity}
C.~Papadimitriou.
\newblock {\em Computational Complexity}.
\newblock Addison-Wesley, 1994.

\bibitem[Rog67]{rog:b:rft}
H.~Rogers, Jr.
\newblock {\em The Theory of Recursive Functions and Effective Computability}.
\newblock McGraw-Hill, 1967.

\bibitem[Sch76]{sch:c:self-reducible}
C.~Schnorr.
\newblock Optimal algorithms for self-reducible problems\typeout{MINOR PANIC:
  Missing Publisher/LNCS number if it was LNCS at all}.
\newblock In {\em Proceedings of the 3rd International Colloquium on Automata,
  Languages, and Programming}, pages 322--337, July 1976.

\bibitem[Sel82]{sel:j:reductions-pselective}
A.~Selman.
\newblock Reductions on {N}{P} and {P}-selective sets.
\newblock {\em Theoretical Computer Science}, 19(3):287--304, 1982.

\end{thebibliography}

\end{document}